\documentclass[aps,prl,twocolumn,amsmath,amssymb]{revtex4-1}

\usepackage{amsmath}
\usepackage{amstext}
\usepackage{bm}
\usepackage{xcolor}
\usepackage{graphicx}

\usepackage[unicode=true,colorlinks=true]{hyperref}

\begin{document}

\title{Strongly temperature dependent resistance of meander-patterned graphene}
\author{G. Yu. Vasileva}
\affiliation{Institut f{\"u}r Festk{\"o}rperphysik, Universit{\"a}t Hannover, Hannover, 30167 Germany}
\affiliation{Ioffe Institute, 194021 St. Petersburg, Russia}
\author{D. Smirnov}
\affiliation{Institut f{\"u}r Festk{\"o}rperphysik, Universit{\"a}t Hannover, Hannover, 30167 Germany}
\author{Yu. B. Vasilyev}
\affiliation{Ioffe Institute, 194021 St. Petersburg, Russia}
\author{M.O.~Nestoklon}
\affiliation{Ioffe Institute, 194021 St. Petersburg, Russia}
\author{N. S. Averkiev}
\affiliation{Ioffe Institute, 194021 St. Petersburg, Russia}
\author{S. Novikov}
\affiliation{Micro and Nanoscience Laboratory, Aalto University, Tietotie 3, FIN-02150, Espoo, Finland}
\author{I. I. Kaya}
\affiliation{Sabanci University,Faculty of Engineering and Natural Sciences, 34956, Istanbul, Turkey}
\author{R. J. Haug}
\affiliation{Institut f{\"u}r Festk{\"o}rperphysik, Universit{\"a}t Hannover, Hannover, 30167 Germany}


\begin{abstract}
We have studied the electronic properties of epitaxial graphene devices patterned in a meander shape with the length up to a few centimeters and the width of few tens of microns. These samples show a pronounced dependence of the resistance on temperature. Accurate comparison with theory shows that this temperature dependence originates from the weak localization effect observed over a broad temperature range from 1.5~K up to 77~K. The comparison allows us to estimate the characteristic times related to quantum interference. In addition, a large resistance enhancement with temperature is observed at the quantum Hall regime near the filling factor of 2. Record high resistance and its strong temperature dependence are favorable for the construction of bolometric photodetectors.
\end{abstract}

\maketitle


Due to its unique properties, graphene is a promising material for constructing emitters and detectors of different frequency, including poorly developed terahertz (THz) range. THz hot electron bolometers, operating at cryogenic temperatures, were made of monolayer\cite{Fang12, Vicarelli12, McKitteick13} and bilayer graphene.\cite{Yan12, Kim13} However, the resistance of graphene typically depends weakly on the electron temperature, which is an obstacle to the creation of graphene-based devices with high efficiency. Recently, successful observation of bolometric signal in graphene was achieved in highly resistive samples of two different types: either in a strongly disordered graphene\cite{Han13} or in a patterned sample.\cite{Vasilyev14, Fatimy15} In particular, epitaxial graphene on SiC patterned in a meander shape has been reported to demonstrate good performance in the terahertz region (1-3 THz).\cite{Vasilyev14}

In this paper we present magnetotransport studies of meander shaped samples with the ratio of the sample length to width  exceeding 600. We show that without or at weak magnetic fields strong temperature dependence  of the resistance up to 70~K is related to weak localization effect.\cite{Hikami80,Kawabata84,McCann06,Nestoklon11,Nestoklon13,Nestoklon14} We investigate this effect in a wide range of temperatures and magnetic fields and find that our data are perfectly fitted by theory.\cite{McCann06} The times of dephasing, intra-valley and inter-valley scattering are obtained from the fitting. Another area of strong temperature dependence of the resistance is found in the quantum Hall regime at magnetic fields with the filling factor being equal to 2, where the longitudinal resistance is zero. The observed large temperature coefficient of the resistance in graphene near the integer QHE may be used to realize a photoresponse mechanisms related to optically induced breakdown of the quantum Hall effect.\cite{Vasilev92} These effects were investigated and confirmed in semiconductor heterostructures, but, until now, studies of graphene samples with a large ratio of length to width $L / W$ have not been carried out.

\begin{figure}
\includegraphics[width=\linewidth]{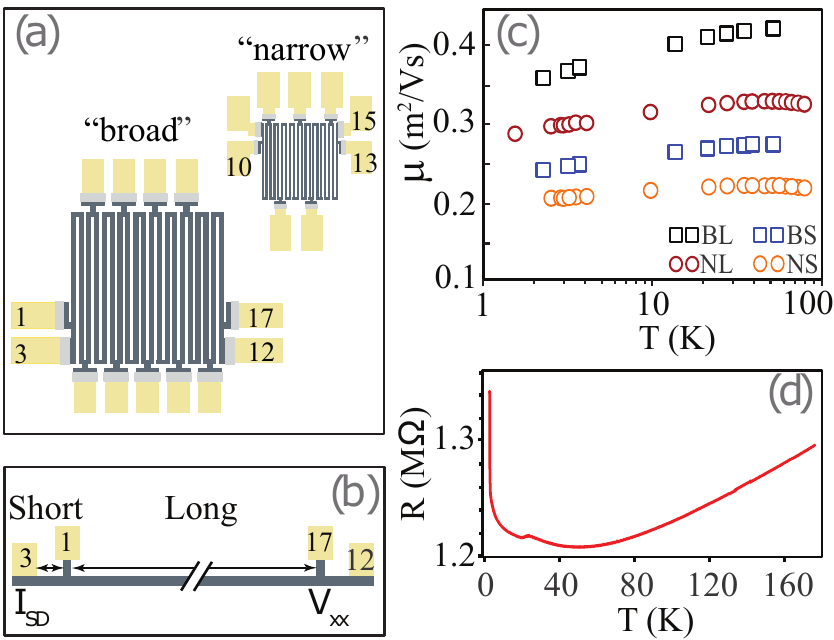}
\caption{Schematic view of the samples. a)  A meander-shaped long 2DEG Hall bars with two different widths, b) An equivalent geometry of the samples with a long narrow stripe. c) Mobility obtained in transport measurements for different samples. d) Typical temperature dependence of the resistance.
}\label{fig:sample}
\end{figure}

The graphene samples were obtained by sublimation at the Si face of a 4H-SiC substrate by thermal decomposition in an argon atmosphere. Raman spectroscopy shows that graphene is a monolayer with high quality. Using laser lithography, samples were patterned in meander shaped Hall devices as shown in Fig.~\ref{fig:sample}(a). Measurements were performed on two samples: "broad" (B) has a width of 50~$\mu$m and "narrow" (N) - 25~$\mu$m. Each sample has several Ti/Au 5/50~nm contacts which were made by e-beam evaporation and lift-off photolithography. Measurements were performed for "short" (S) neighboring contacts which are close to each other (for example 1-3 or 15-13) and for "long" (L) contact with long distance between them (for example 1-17 or 10-13). Because of significant length-to-width ratio difference between L contact ($\sim$610) and S contact ($\sim$2), we consider the results obtained for different contacts as the results for different samples. An equivalent scheme of the meander-structure is depicted in Fig.~\ref{fig:sample}(b), for better understanding of the sample geometry. Table~\ref{tbl:samples} compiles the parameters of the samples at T=2~K. The resistance was measured by lock-in amplifiers with low-frequency currents (f = 17.7 Hz, I = 100 nA) in a temperature range 1.5-200~K and at magnetic fields up to 12T.

\begin{table}
\begin{tabular}{|c|ccccc|}
\hline
Sample N    & $W$($\mu$m) & $L$($10^{3}\mu$m) & $n$($10^{11}$cm$^{-2}$)& $\mu$(cm$^2$/Vs) & L/W \\ \hline
BS  &   50  &  0.1   &  5.8   &   2300    &   2  \\
BL  &    50 &  30.5   &  4.6   &    3500   &  610   \\
NS & 25    & 0.05    &  8.5   &   2000    &  2   \\
NL & 25    & 15.25     &  6.8   &   2800    &  610   \\ \hline
\end{tabular}
\caption{Parameters of the samples.}\label{tbl:samples}
\end{table}


The electron densities and the mobilities determined at liquid helium temperatures are presented in Table~\ref{tbl:samples}. The mobility was calculated from zero magnetic field conductivity as $\mu=\sigma/ne$ and it demonstrates a logarithmic dependence on temperature in a wide range as shown in Fig.~\ref{fig:sample}(c). Such logarithmic behavior is typical for the weak localization effect which will be discussed below. One can see that the mobility is lower between S contacts in comparison with L contacts in the same sample. This is correlated with the fact that the electron density is by 24\% higher in S samples than in the L samples \cite{mobden}. The density has been extracted from Shubnikov-de Haas oscillations (SdHO) at 1.5~K under the assumption of a homogeneous electron distribution in the sample. For a S samples magnetoresistance (MR) oscillations are resolved better than for a L samples. At the same time, the position of the maxima and minima of the MR oscillations are shifted with respect to each other in the magnetic field which means different charge carrier concentrations in these samples. This effect is probably associated with inhomogeneous charge distributions in graphene samples caused, in particular, by surface ripples.\cite{DasSarma11, Juan07} In reality, due to disorder, the electron distribution is inhomogeneous along the plane of graphene. Therefore, these values should not be considered as the real electron densities, but as some effective carrier densities. The difference in electron concentration cannot be explained by uncertainty in the sample size because the latter was obtained from the magnetoresistance oscillations. We suppose that the strong inhomogeneity of disorder in the samples is responsible for this fact.

Figure~\ref{fig:sample}(d) shows the resistance as a function of T for sample BL. One can see that the resistance decreases with decreasing temperature in the range from 200~K down to 50~K. Afterwards it increases for further decreasing temperature and finally diverges at low temperatures (as shown in Fig.~\ref{fig:sample}(d)). There is also some peculiarity in the dependence around $T=25$~K. The complicated temperature dependence may be attributed to different scattering mechanisms dominating at different temperature ranges. The strongest variation of electrical resistance with temperature is clearly visible below 6~K with a temperature coefficient for the resistance as high as 40~kOhm/K. To reveal the origin of this temperature dependence additional magnetotransport measurements were done.

\begin{figure}
	\includegraphics[width=\linewidth]{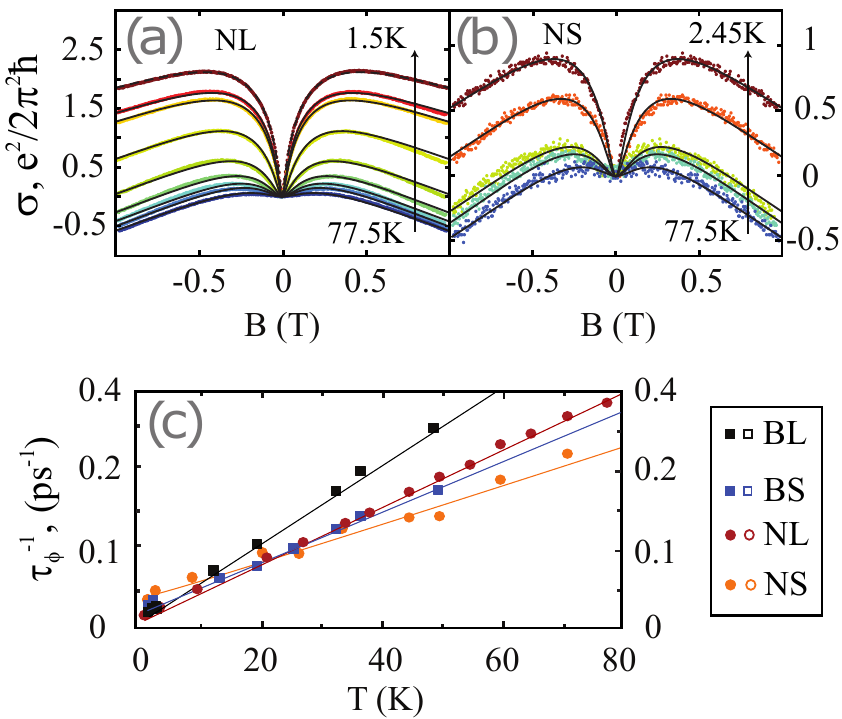}
	\caption{The normalized magnetoconductance at weak magnetic fields for narrow sample for (a) long contacts at $T=1.5;2.45;3.9;9.5;21;34;45;55;77.5~K$ and (b) short contacts at $T=2.45;3.5;50;77.5~K$.
		Thin black lines represent the best fit using Equation~\ref{eq:HLN}, color dots are experimental data.
		(c) Inverse dephasing time shows linear temperature dependence for all samples. Color symbols-experimental data and solid line-linear fit.
	}\label{fig:WL}
\end{figure}

The longitudinal resistance $R_{xx}$ was measured as a function of applied perpendicular field $B$ in more details in the range from $-1$ to $1$~T. To exclude a contribution of Hall resistance for the sample BS with short length arising due to asymmetric contacts we consider below the symmetrical part of the resistance $(R_{xx}(B) + R_{xx}(-B))$. The further analysis of the experimental data was performed with the longitudinal conductivity. A series of magnetoconductance traces for a broad range of temperature between 1.5 and 77~K are depicted for the samples NL and NS in Fig.~\ref{fig:WL}(a) and (b), respectively. Experimental data are shown in dots of different colors and solid curves are the results of calculation as best fit to experiments (see below). The sharp positive magnetoconductance around zero magnetic fields shown in Fig.~\ref{fig:WL} is qualitatively rather common for the WL effect except the very strong value of WL magnitude reaching 1~MOhm. It is seen that even at temperatures of 77~K WL is still preserved. 

These results of the transport measurements around $B=0$ can be explained by the weak localization theory which takes into account contributions of quantum corrections to the conductivity. 
In graphene one can observe both types of quantum corrections: weak localization and weak antilocalization effects depending, in particular, on the dominating type of scattering.\cite{Wu07,Tikhonenko08} Since graphene has two nonequivalent Dirac cones on opposite sides of the Brillouin zone, scattering includes inter-valley transitions. Inter-valley transitions consume a large momentum transfer that can be realized when carriers are scattered by atomically sharp defects, such as the edge of the sample. Inter-valley scattering suppresses the weak antilocalization and simultaneously leads to weak localization.\cite{McCann06,Nestoklon14}


To fit the experimental data in Fig.~\ref{fig:WL}(a),(b) with the analytical results we use the following procedure: First, we extract the Fermi wave vector from the measured carrier concentration:
\begin{equation}
	k_F = \sqrt{\pi n} \;.
\end{equation}
From conductivity at zero magnetic fields one may then extract mean free path 
length $\ell$ (and time associated with this length 
$\tau = \ell/v_F$ where Fermi velocity $v_F=10^8$cm/s in graphene 
does not depend on Fermi energy) using Drude conductivity 
in the form
\begin{equation}
	\sigma_{D} = \frac4{R_K} k_F\ell  \;.
\end{equation}
Here we use von Klitzing constant $R_K = h/e^2 = 25812.8$~Ohm

The value of the mean free path length allows us to define transport magnetic field 
\begin{equation}
	B_{\text{tr}} =  \frac{\hbar}{2e \ell^2} \;. 
\end{equation}

Hikami-Larkin-Nagaoka theory\cite{Hikami80} in graphene leads to a conductivity correction 
caused by weak localization which is controlled by three characteristic
times: $\tau_{\phi}$ is dephasing time; $\tau_i$ is intervalley scattering time and 
$\tau_*$ is the time associated with elastic intravalley scattering with reduced symmetry.\cite{McCann06}
We rewrite the conductivity correction as
\begin{multline}\label{eq:HLN}
	\Delta \sigma = \frac1{\pi R_K} \left[ 
	F_2\left( \frac{\tau_{\phi}}{\tau} \frac{B}{B_{\text{tr}}} \right) - 
	\right. \\ \left.
	F_2\left( \frac1{ \left(\frac{\tau_{\phi}}{\tau} \right)^{-1} +
		2 \left(\frac{\tau_{i}}{\tau}\right)^{-1} }  \frac{B}{B_{\text{tr}}} \right) -
	\right. \\ \left.
	2 F_2\left( \frac1 { \left(\frac{\tau_{\phi}}{\tau} \right)^{-1} +
		\left(\frac{\tau_{i}}{\tau} \right)^{-1} +
		\left(\frac{\tau_{*}}{\tau} \right)^{-1} }
	\frac{B}{B_{\text{tr}}} \right)
	\right],
\end{multline}
where we use auxiliary function
\begin{equation}
	F_2(x) = \psi\left( \frac12 + \frac1x \right) + \log(x) \;,
\end{equation}
here $\psi(y)$ is digamma function.

We use least square procedure varying $\tau_{\phi}/\tau$ , $\tau_i/\tau$ and $\tau_*/\tau$ to find the best 
fit of the analytic result \eqref{eq:HLN} to experimental data.
It should be noted however that the procedure is ambiguous and allows for two variants: 
one is $\tau_*\gg\tau_i$ another is $\tau_*\sim \tau_i$. 
In the minimization, we assume that $\tau_* \gg \tau_i$ because otherwise the 
equality $\tau_*=\tau_i$ is dictated by minimization procedure.
Previously it has been reported\cite{Tikhonenko08,Ki08} that in some samples 
the case $\tau_*\ll\tau_i$ is realized. While to distinguish between 
$\tau_*\gg\tau_i$ and $\tau_*\sim \tau_i$ cases we need more data, 
the case $\tau_*\ll\tau_i$ is not consistent with our experimental data.

Such an analysis of the experimental date yielded the change of scattering times with the temperature. The intra-valley scattering time $\tau$ and inter-valley scattering time $\tau_{*}$ show only a weak  temperature dependence. The dephasing time $\tau_{\phi}$, which controls the magnitude of the weak localization effect, strongly decreases with temperature. This explains, why the WL effect is destroyed by increasing temperature. More detailed analysis demonstrates that $1/\tau_{\phi}$ is a linear function of $T$ as expected for electron-electron scattering in the diffusive regime. These data together with the linear fitting are depicted in Fig~\ref{fig:WL}(c).

The estimated decoherence time can reach 10.2 ps for the sample BL and 3.7 ps for the sample BS at the lowest temperatures, and it decreases with increasing temperature. These values are less than the 50 ps established as a limit of spin relaxation time.\cite{LaraAvila11} This fact together with the linear temperature dependence of the decoherence rate excludes the importance of magnetic impurities in our samples. Most likely, the decoherence rate is defined by scattering on impurities on the SiC surface \cite{Huang15} which is faster than intervalley scattering at low temperatures. Comparing the scattering times measured in our samples with data found in other epitaxial graphene samples we find that they are consistent with the previously reported results.\cite{Tanabe11,Giesbers13,Yu13} This finding suggests that our samples have good quality. However, they contain some amounts of disorder, because the actual characteristics vary from sample to sample. Which is more important, the lithography processing has negligible effect on the sample quality. The long samples with a large edge length demonstrate clearly a temperature effect on resistance which is the same as in the short samples. In the WL effect, the relation of the relative resistance change with temperature does not depend strongly on the sample shape.
At the same time, the increase of the sample length leads to increase of the sample resistance considerably and in turn, as expected, could improve the photodetectivity and responsivity of such samples.
This finding demonstrates that patterned graphene structures with high resistances formed in specific shapes can be used for various types of photodetectors.

\begin{figure}
	\includegraphics[width=\linewidth]{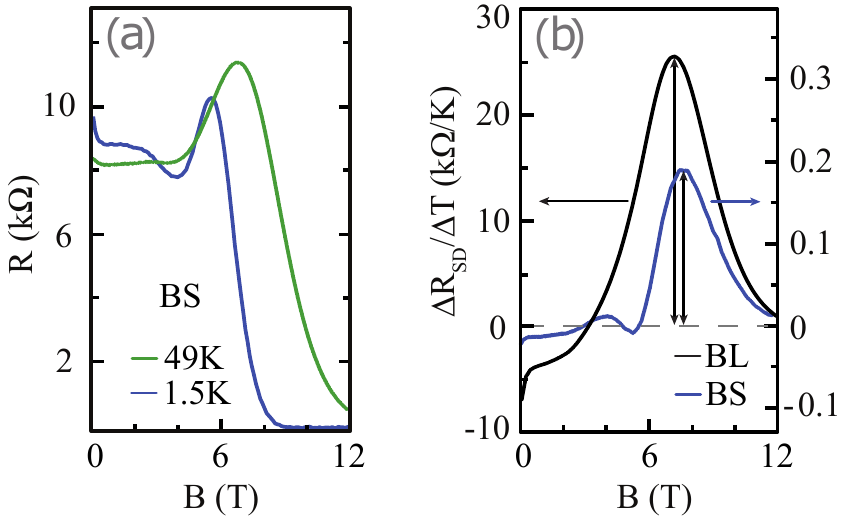}
	\caption{a) Shubnikov-de Haas oscillations for BS sample at 1.5K (blue line) and 49~K (green line). b) The change of the resistance with temperature in magnetic field for BL (black line) and BS (blue line) samples was calculated from magnetoresistance.
	}\label{fig:SdHO}
\end{figure}

Resistance variations with temperature were also found at higher magnetic fields. Magnetoresistance traces were recorded at different temperatures. Fig.~\ref{fig:SdHO}(a) shows the resistance as a function of magnetic field from $0$ to $12$~T, at $1.5$ and $49$~K for the BS sample. The resistance demonstrates nonmonotonic behaviour with a drop near the zero magnetic field followed by smooth oscillations. From these data we are able to estimate the change of the resistance with temperature shown in Fig.~\ref{fig:SdHO}(b) for BS and BL samples, calculated for resistance measured at T= 1.5 and 49 K. The data in Fig.~\ref{fig:SdHO}(b) gives the values which can be considered as the characteristic of the bolometric responsivity of the sample, as was shown in the work \cite{Vasilev92} which reported the photo-induced breakdown of the QHE. One can compare this value ($\sim 25600$~Ohm/K) with the value obtained from the photoresponse in an ordinary graphene sample where it is 18~Ohm/K (Fig.~3 in Ref.~\onlinecite{Kalugin11}). The higher this value, the larger is the expected photoconductivity signal. From this point of view, the data for the sample BL is very notable. The change of the resistance with temperature is controlled by the ratio of the sample length to its width. In BL sample the strong effect is obtained due to the large ratio of length to width in comparison to conventional Hall bars, which is, 2-5 typically for graphene samples.

These transport measurements allow us to estimate the prospect of structured graphene as QHE bolometric detectors.\cite{Vasilev92}  At the integer QHE in two-dimensional electron system, an integer number of Landau level is completely filled with electrons. The longitudinal resistance $R_{xx}$ of such a system vanishes, and the Hall resistance $R_H$ is quantized. Illumination by light with a photon energy equal to cyclotron energy leads to a nonzero longitudinal resistance, i.e., photoconductivity which is caused by the dissipation of electrons being redistributed between the Landau levels under resonant radiation. The magnitude of the photoresponse is proportional to the ratio of the sample length to width $L / W$ \cite{Vas2002}. Therefore, an increase of the ratio $L / W$ leads to an increase in the photoresponse signal ceteris paribus. A remarkable feature of the photodetectors in the integer QHE regime is the absence of the dissipative resistance $R_{xx} = 0$, so that with increasing $L / W$ there is no increase of the background noise, which depends on the resistance of the sample. In Ref.~\onlinecite{Kalugin11} photoconductivity signal was found near the flake of the QHE plateau exactly where the temperature coefficient of the resistance reaches a maximum. Note that in this ref. the maximum is 1000 times less then for BL sample. This gives us a reason to expect that such remarkably strong temperature sensitivity of the resistance in meander samples should lead to the strong photoresponse. Moreover without or at low magnetic field the meander shape samples demonstrate strong temperature dependents due to WL. We predict that in THz and IR spectral ranges it is possible to use two different effects: the weak localization and the quantum Hall effect to create highly sensitive detectors.
'


In summary, we have studied the electronic properties of epitaxial graphene samples patterned in a meander-shaped Hall devices with lengths up to a few centimeters and widths of only few tens of microns. These samples show a pronounced low-temperature strong resistance derivative as a result of weak localization. Weak localization in the patterned graphene samples is observed over a broad temperature range from 1.5 K up to 77 K. The magnetic field dependence of the transport caused by the weak localization is perfectly fitted by theory, which allows us to estimate the characteristic scattering times. In addition, a large resistance enhancement with temperature is observed at the quantum Hall regime near the filling factor 2. 
Our work demonstrates that patterned graphene with high resistance could be used as THz bolometer due to strong temperature dependence of its resistance. In particular, WL and QH effects are found to be effective for this purpose in meander-patterned graphene devices.


G.Yu.V. and Yu.B.V. are grateful to I.V. Gornyi,  A.P. Dmitriev and V.Yu. Kachorovskii for helpful discussions. This work was supported by RAN and RFBR. MN acknowledges financial support from RFBR grant 16-02-00375


\bibliography{Transport_15}

\begin{thebibliography}{28}%
\makeatletter
\providecommand \@ifxundefined [1]{%
 \@ifx{#1\undefined}
}%
\providecommand \@ifnum [1]{%
 \ifnum #1\expandafter \@firstoftwo
 \else \expandafter \@secondoftwo
 \fi
}%
\providecommand \@ifx [1]{%
 \ifx #1\expandafter \@firstoftwo
 \else \expandafter \@secondoftwo
 \fi
}%
\providecommand \natexlab [1]{#1}%
\providecommand \enquote  [1]{``#1''}%
\providecommand \bibnamefont  [1]{#1}%
\providecommand \bibfnamefont [1]{#1}%
\providecommand \citenamefont [1]{#1}%
\providecommand \href@noop [0]{\@secondoftwo}%
\providecommand \href [0]{\begingroup \@sanitize@url \@href}%
\providecommand \@href[1]{\@@startlink{#1}\@@href}%
\providecommand \@@href[1]{\endgroup#1\@@endlink}%
\providecommand \@sanitize@url [0]{\catcode `\\12\catcode `\$12\catcode
  `\&12\catcode `\#12\catcode `\^12\catcode `\_12\catcode `\%12\relax}%
\providecommand \@@startlink[1]{}%
\providecommand \@@endlink[0]{}%
\providecommand \url  [0]{\begingroup\@sanitize@url \@url }%
\providecommand \@url [1]{\endgroup\@href {#1}{\urlprefix }}%
\providecommand \urlprefix  [0]{URL }%
\providecommand \Eprint [0]{\href }%
\providecommand \doibase [0]{http://dx.doi.org/}%
\providecommand \selectlanguage [0]{\@gobble}%
\providecommand \bibinfo  [0]{\@secondoftwo}%
\providecommand \bibfield  [0]{\@secondoftwo}%
\providecommand \translation [1]{[#1]}%
\providecommand \BibitemOpen [0]{}%
\providecommand \bibitemStop [0]{}%
\providecommand \bibitemNoStop [0]{.\EOS\space}%
\providecommand \EOS [0]{\spacefactor3000\relax}%
\providecommand \BibitemShut  [1]{\csname bibitem#1\endcsname}%
\let\auto@bib@innerbib\@empty
\bibitem [{\citenamefont {Fang}\ \emph {et~al.}(2012)\citenamefont {Fang},
  \citenamefont {Liu}, \citenamefont {Wang}, \citenamefont {Ajayan},
  \citenamefont {Nordlander},\ and\ \citenamefont {Halas}}]{Fang12}%
  \BibitemOpen
  \bibfield  {author} {\bibinfo {author} {\bibfnamefont {Z.}~\bibnamefont
  {Fang}}, \bibinfo {author} {\bibfnamefont {Z.}~\bibnamefont {Liu}}, \bibinfo
  {author} {\bibfnamefont {Y.}~\bibnamefont {Wang}}, \bibinfo {author}
  {\bibfnamefont {P.~M.}\ \bibnamefont {Ajayan}}, \bibinfo {author}
  {\bibfnamefont {P.}~\bibnamefont {Nordlander}}, \ and\ \bibinfo {author}
  {\bibfnamefont {N.~J.}\ \bibnamefont {Halas}},\ }\href
  {http://dx.doi.org/10.1021/nl301774e} {\bibfield  {journal} {\bibinfo
  {journal} {Nano Letters}\ }\textbf {\bibinfo {volume} {12}},\ \bibinfo
  {pages} {3808} (\bibinfo {year} {2012})}\BibitemShut {NoStop}%
\bibitem [{\citenamefont {Vicarelli}\ \emph {et~al.}(2012)\citenamefont
  {Vicarelli}, \citenamefont {Vitiello}, \citenamefont {Coquillat},
  \citenamefont {Lombardo}, \citenamefont {Ferrari}, \citenamefont {Knap},
  \citenamefont {Polini}, \citenamefont {Pellegrini},\ and\ \citenamefont
  {Tredicucci}}]{Vicarelli12}%
  \BibitemOpen
  \bibfield  {author} {\bibinfo {author} {\bibfnamefont {L.}~\bibnamefont
  {Vicarelli}}, \bibinfo {author} {\bibfnamefont {M.~S.}\ \bibnamefont
  {Vitiello}}, \bibinfo {author} {\bibfnamefont {D.}~\bibnamefont {Coquillat}},
  \bibinfo {author} {\bibfnamefont {A.}~\bibnamefont {Lombardo}}, \bibinfo
  {author} {\bibfnamefont {A.~C.}\ \bibnamefont {Ferrari}}, \bibinfo {author}
  {\bibfnamefont {W.}~\bibnamefont {Knap}}, \bibinfo {author} {\bibfnamefont
  {M.}~\bibnamefont {Polini}}, \bibinfo {author} {\bibfnamefont
  {V.}~\bibnamefont {Pellegrini}}, \ and\ \bibinfo {author} {\bibfnamefont
  {A.}~\bibnamefont {Tredicucci}},\ }\href
  {http://www.nature.com/nmat/journal/v11/n10/full/nmat3417.html} {\bibfield
  {journal} {\bibinfo  {journal} {Nature Materials}\ }\textbf {\bibinfo
  {volume} {11}},\ \bibinfo {pages} {865} (\bibinfo {year} {2012})}\BibitemShut
  {NoStop}%
\bibitem [{\citenamefont {McKitterick}, \citenamefont {Prober},\ and\
  \citenamefont {Karasik}(2013)}]{McKitteick13}%
  \BibitemOpen
  \bibfield  {author} {\bibinfo {author} {\bibfnamefont {C.~B.}\ \bibnamefont
  {McKitterick}}, \bibinfo {author} {\bibfnamefont {D.~E.}\ \bibnamefont
  {Prober}}, \ and\ \bibinfo {author} {\bibfnamefont {B.~S.}\ \bibnamefont
  {Karasik}},\ }\href
  {http://scitation.aip.org/content/aip/journal/jap/113/4/10.1063/1.4789360}
  {\bibfield  {journal} {\bibinfo  {journal} {Journal of Applied Physics}\
  }\textbf {\bibinfo {volume} {113}},\ \bibinfo {eid} {044512} (\bibinfo {year}
  {2013})}\BibitemShut {NoStop}%
\bibitem [{\citenamefont {Yan}\ \emph {et~al.}(2012)\citenamefont {Yan},
  \citenamefont {Kim}, \citenamefont {Elle}, \citenamefont {Sushkov},
  \citenamefont {Jenkins}, \citenamefont {Milchberg}, \citenamefont {Fuhrer},\
  and\ \citenamefont {Drew}}]{Yan12}%
  \BibitemOpen
  \bibfield  {author} {\bibinfo {author} {\bibfnamefont {J.}~\bibnamefont
  {Yan}}, \bibinfo {author} {\bibfnamefont {M.-H.}\ \bibnamefont {Kim}},
  \bibinfo {author} {\bibfnamefont {J.~A.}\ \bibnamefont {Elle}}, \bibinfo
  {author} {\bibfnamefont {A.~B.}\ \bibnamefont {Sushkov}}, \bibinfo {author}
  {\bibfnamefont {G.~S.}\ \bibnamefont {Jenkins}}, \bibinfo {author}
  {\bibfnamefont {H.~M.}\ \bibnamefont {Milchberg}}, \bibinfo {author}
  {\bibfnamefont {M.~S.}\ \bibnamefont {Fuhrer}}, \ and\ \bibinfo {author}
  {\bibfnamefont {H.~D.}\ \bibnamefont {Drew}},\ }\href
  {http://www.nature.com/nnano/journal/v7/n7/full/nnano.2012.88.html}
  {\bibfield  {journal} {\bibinfo  {journal} {Nature Nanotechnology}\ }\textbf
  {\bibinfo {volume} {7}},\ \bibinfo {pages} {472} (\bibinfo {year}
  {2012})}\BibitemShut {NoStop}%
\bibitem [{\citenamefont {Kim}\ \emph {et~al.}(2013)\citenamefont {Kim},
  \citenamefont {Yan}, \citenamefont {Suess}, \citenamefont {Murphy},
  \citenamefont {Fuhrer},\ and\ \citenamefont {Drew}}]{Kim13}%
  \BibitemOpen
  \bibfield  {author} {\bibinfo {author} {\bibfnamefont {M.-H.}\ \bibnamefont
  {Kim}}, \bibinfo {author} {\bibfnamefont {J.}~\bibnamefont {Yan}}, \bibinfo
  {author} {\bibfnamefont {R.~J.}\ \bibnamefont {Suess}}, \bibinfo {author}
  {\bibfnamefont {T.~E.}\ \bibnamefont {Murphy}}, \bibinfo {author}
  {\bibfnamefont {M.~S.}\ \bibnamefont {Fuhrer}}, \ and\ \bibinfo {author}
  {\bibfnamefont {H.~D.}\ \bibnamefont {Drew}},\ }\href
  {http://link.aps.org/doi/10.1103/PhysRevLett.110.247402} {\bibfield
  {journal} {\bibinfo  {journal} {Phys. Rev. Lett.}\ }\textbf {\bibinfo
  {volume} {110}},\ \bibinfo {pages} {247402} (\bibinfo {year}
  {2013})}\BibitemShut {NoStop}%
\bibitem [{\citenamefont {Han}\ \emph {et~al.}(2013)\citenamefont {Han},
  \citenamefont {Gao}, \citenamefont {Zhang}, \citenamefont {Chen},
  \citenamefont {Chen}, \citenamefont {Liu}, \citenamefont {Zhang},
  \citenamefont {Liu}, \citenamefont {Wu}, ,\ and\ \citenamefont {Yu}}]{Han13}%
  \BibitemOpen
  \bibfield  {author} {\bibinfo {author} {\bibfnamefont {Q.}~\bibnamefont
  {Han}}, \bibinfo {author} {\bibfnamefont {T.}~\bibnamefont {Gao}}, \bibinfo
  {author} {\bibfnamefont {R.}~\bibnamefont {Zhang}}, \bibinfo {author}
  {\bibfnamefont {Y.}~\bibnamefont {Chen}}, \bibinfo {author} {\bibfnamefont
  {J.}~\bibnamefont {Chen}}, \bibinfo {author} {\bibfnamefont {G.}~\bibnamefont
  {Liu}}, \bibinfo {author} {\bibfnamefont {Y.}~\bibnamefont {Zhang}}, \bibinfo
  {author} {\bibfnamefont {Z.}~\bibnamefont {Liu}}, \bibinfo {author}
  {\bibfnamefont {X.}~\bibnamefont {Wu}}, , \ and\ \bibinfo {author}
  {\bibfnamefont {D.}~\bibnamefont {Yu}},\ }\href
  {http://www.ncbi.nlm.nih.gov/pmc/articles/PMC3866633/} {\bibfield  {journal}
  {\bibinfo  {journal} {Scientific Reports}\ ,\ \bibinfo {pages} {3533}}
  (\bibinfo {year} {2013})}\BibitemShut {NoStop}%
\bibitem [{\citenamefont {Vasilyev}\ \emph {et~al.}(2014)\citenamefont
  {Vasilyev}, \citenamefont {Vasileva}, \citenamefont {Ivanov}, \citenamefont
  {Novikov},\ and\ \citenamefont {Danilov}}]{Vasilyev14}%
  \BibitemOpen
  \bibfield  {author} {\bibinfo {author} {\bibfnamefont {Y.~B.}\ \bibnamefont
  {Vasilyev}}, \bibinfo {author} {\bibfnamefont {G.~Y.}\ \bibnamefont
  {Vasileva}}, \bibinfo {author} {\bibfnamefont {Y.~L.}\ \bibnamefont
  {Ivanov}}, \bibinfo {author} {\bibfnamefont {S.}~\bibnamefont {Novikov}}, \
  and\ \bibinfo {author} {\bibfnamefont {S.~N.}\ \bibnamefont {Danilov}},\
  }\href
  {http://scitation.aip.org/content/aip/journal/apl/105/17/10.1063/1.4900788}
  {\bibfield  {journal} {\bibinfo  {journal} {Applied Physics Letters}\
  }\textbf {\bibinfo {volume} {105}},\ \bibinfo {eid} {171105} (\bibinfo {year}
  {2014})}\BibitemShut {NoStop}%
\bibitem [{\citenamefont {Fatimy}\ \emph {et~al.}(2015)\citenamefont {Fatimy},
  \citenamefont {Myers-Ward}, \citenamefont {Boyd}, \citenamefont {Daniels},
  \citenamefont {Gaskill},\ and\ \citenamefont {Barbara}}]{Fatimy15}%
  \BibitemOpen
  \bibfield  {author} {\bibinfo {author} {\bibfnamefont {A.~E.}\ \bibnamefont
  {Fatimy}}, \bibinfo {author} {\bibfnamefont {R.~L.}\ \bibnamefont
  {Myers-Ward}}, \bibinfo {author} {\bibfnamefont {A.~K.}\ \bibnamefont
  {Boyd}}, \bibinfo {author} {\bibfnamefont {K.~M.}\ \bibnamefont {Daniels}},
  \bibinfo {author} {\bibfnamefont {D.~K.}\ \bibnamefont {Gaskill}}, \ and\
  \bibinfo {author} {\bibfnamefont {P.}~\bibnamefont {Barbara}},\ }\href
  {http://www.nature.com/nnano/journal/v11/n4/full/nnano.2015.303.html}
  {\bibfield  {journal} {\bibinfo  {journal} {Nature Nanotechnology}\ }\textbf
  {\bibinfo {volume} {11}},\ \bibinfo {pages} {335} (\bibinfo {year}
  {2015})}\BibitemShut {NoStop}%
\bibitem [{\citenamefont {Hikami}, \citenamefont {Larkin},\ and\ \citenamefont
  {Nagaoka}(1980)}]{Hikami80}%
  \BibitemOpen
  \bibfield  {author} {\bibinfo {author} {\bibfnamefont {S.}~\bibnamefont
  {Hikami}}, \bibinfo {author} {\bibfnamefont {A.~I.}\ \bibnamefont {Larkin}},
  \ and\ \bibinfo {author} {\bibfnamefont {Y.}~\bibnamefont {Nagaoka}},\ }\href
  {http://dx.doi.org/10.1143/PTP.63.707} {\bibfield  {journal} {\bibinfo
  {journal} {Prog. Theor. Phys.}\ }\textbf {\bibinfo {volume} {63}},\ \bibinfo
  {pages} {707} (\bibinfo {year} {1980})}\BibitemShut {NoStop}%
\bibitem [{\citenamefont {Kawabata}(1984)}]{Kawabata84}%
  \BibitemOpen
  \bibfield  {author} {\bibinfo {author} {\bibfnamefont {A.}~\bibnamefont
  {Kawabata}},\ }\href {http://dx.doi.org/10.1143/JPSJ.53.3540} {\bibfield
  {journal} {\bibinfo  {journal} {Journal of the Physical Society of Japan}\
  }\textbf {\bibinfo {volume} {53}},\ \bibinfo {pages} {3540} (\bibinfo {year}
  {1984})}\BibitemShut {NoStop}%
\bibitem [{\citenamefont {McCann}\ \emph {et~al.}(2006)\citenamefont {McCann},
  \citenamefont {Kechedzhi}, \citenamefont {Fal'ko}, \citenamefont {Suzuura},
  \citenamefont {Ando},\ and\ \citenamefont {Altshuler}}]{McCann06}%
  \BibitemOpen
  \bibfield  {author} {\bibinfo {author} {\bibfnamefont {E.}~\bibnamefont
  {McCann}}, \bibinfo {author} {\bibfnamefont {K.}~\bibnamefont {Kechedzhi}},
  \bibinfo {author} {\bibfnamefont {V.~I.}\ \bibnamefont {Fal'ko}}, \bibinfo
  {author} {\bibfnamefont {H.}~\bibnamefont {Suzuura}}, \bibinfo {author}
  {\bibfnamefont {T.}~\bibnamefont {Ando}}, \ and\ \bibinfo {author}
  {\bibfnamefont {B.~L.}\ \bibnamefont {Altshuler}},\ }\href
  {http://link.aps.org/doi/10.1103/PhysRevLett.97.146805} {\bibfield  {journal}
  {\bibinfo  {journal} {Phys. Rev. Lett.}\ }\textbf {\bibinfo {volume} {97}},\
  \bibinfo {pages} {146805} (\bibinfo {year} {2006})}\BibitemShut {NoStop}%
\bibitem [{\citenamefont {Nestoklon}, \citenamefont {Averkiev},\ and\
  \citenamefont {Tarasenko}(2011)}]{Nestoklon11}%
  \BibitemOpen
  \bibfield  {author} {\bibinfo {author} {\bibfnamefont {M.~O.}\ \bibnamefont
  {Nestoklon}}, \bibinfo {author} {\bibfnamefont {N.~S.}\ \bibnamefont
  {Averkiev}}, \ and\ \bibinfo {author} {\bibfnamefont {S.~A.}\ \bibnamefont
  {Tarasenko}},\ }\href
  {http://www.sciencedirect.com/science/article/pii/S003810981100384X}
  {\bibfield  {journal} {\bibinfo  {journal} {Solid State Communications}\
  }\textbf {\bibinfo {volume} {151}},\ \bibinfo {pages} {1550} (\bibinfo {year}
  {2011})}\BibitemShut {NoStop}%
\bibitem [{\citenamefont {Nestoklon}\ and\ \citenamefont
  {Averkiev}(2013)}]{Nestoklon13}%
  \BibitemOpen
  \bibfield  {author} {\bibinfo {author} {\bibfnamefont {M.~O.}\ \bibnamefont
  {Nestoklon}}\ and\ \bibinfo {author} {\bibfnamefont {N.~S.}\ \bibnamefont
  {Averkiev}},\ }\href {http://dx.doi.org/10.1209/0295-5075/101/47006}
  {\bibfield  {journal} {\bibinfo  {journal} {EPL (Europhysics Letters)}\
  }\textbf {\bibinfo {volume} {101}},\ \bibinfo {pages} {47006} (\bibinfo
  {year} {2013})}\BibitemShut {NoStop}%
\bibitem [{\citenamefont {Nestoklon}\ and\ \citenamefont
  {Averkiev}(2014)}]{Nestoklon14}%
  \BibitemOpen
  \bibfield  {author} {\bibinfo {author} {\bibfnamefont {M.~O.}\ \bibnamefont
  {Nestoklon}}\ and\ \bibinfo {author} {\bibfnamefont {N.~S.}\ \bibnamefont
  {Averkiev}},\ }\href {http://link.aps.org/doi/10.1103/PhysRevB.90.155412}
  {\bibfield  {journal} {\bibinfo  {journal} {Phys. Rev. B}\ }\textbf {\bibinfo
  {volume} {90}},\ \bibinfo {pages} {155412} (\bibinfo {year}
  {2014})}\BibitemShut {NoStop}%
\bibitem [{\citenamefont {Vasil'ev}\ \emph {et~al.}(1992)\citenamefont
  {Vasil'ev}, \citenamefont {Suchalkin}, \citenamefont {Ivanov}, \citenamefont
  {Ivanov}, \citenamefont {Kop'ev},\ and\ \citenamefont
  {Savel'ev}}]{Vasilev92}%
  \BibitemOpen
  \bibfield  {author} {\bibinfo {author} {\bibfnamefont {Y.~B.}\ \bibnamefont
  {Vasil'ev}}, \bibinfo {author} {\bibfnamefont {S.~D.}\ \bibnamefont
  {Suchalkin}}, \bibinfo {author} {\bibfnamefont {Y.~L.}\ \bibnamefont
  {Ivanov}}, \bibinfo {author} {\bibfnamefont {S.~V.}\ \bibnamefont {Ivanov}},
  \bibinfo {author} {\bibfnamefont {P.~S.}\ \bibnamefont {Kop'ev}}, \ and\
  \bibinfo {author} {\bibfnamefont {I.~G.}\ \bibnamefont {Savel'ev}},\ }\href
  {http://www.jetpletters.ac.ru/ps/1291/article_19505.shtml} {\bibfield
  {journal} {\bibinfo  {journal} {JETP Letters}\ }\textbf {\bibinfo {volume}
  {56}},\ \bibinfo {pages} {377} (\bibinfo {year} {1992})}\BibitemShut
  {NoStop}%
\bibitem [{\citenamefont {Lara-Avila}\ \emph
  {et~al.}(2011{\natexlab{a}})\citenamefont {Lara-Avila}, \citenamefont
  {Moth-Poulsen}, \citenamefont {Yakimova}, \citenamefont {Bjørnholm},
  \citenamefont {Fal’ko}, \citenamefont {Tzalenchuk},\ and\ \citenamefont
  {Kubatkin}}]{mobden}%
  \BibitemOpen
  \bibfield  {author} {\bibinfo {author} {\bibfnamefont {S.}~\bibnamefont
  {Lara-Avila}}, \bibinfo {author} {\bibfnamefont {K.}~\bibnamefont
  {Moth-Poulsen}}, \bibinfo {author} {\bibfnamefont {R.}~\bibnamefont
  {Yakimova}}, \bibinfo {author} {\bibfnamefont {T.}~\bibnamefont
  {Bjørnholm}}, \bibinfo {author} {\bibfnamefont {V.}~\bibnamefont
  {Fal’ko}}, \bibinfo {author} {\bibfnamefont {A.}~\bibnamefont
  {Tzalenchuk}}, \ and\ \bibinfo {author} {\bibfnamefont {S.}~\bibnamefont
  {Kubatkin}},\ }\href {http://dx.doi.org/10.1002/adma.201003993} {\bibfield
  {journal} {\bibinfo  {journal} {Advanced Materials}\ }\textbf {\bibinfo
  {volume} {23}},\ \bibinfo {pages} {878} (\bibinfo {year}
  {2011}{\natexlab{a}})}\BibitemShut {NoStop}%
\bibitem [{\citenamefont {Das~Sarma}\ \emph {et~al.}(2011)\citenamefont
  {Das~Sarma}, \citenamefont {Adam}, \citenamefont {Hwang},\ and\ \citenamefont
  {Rossi}}]{DasSarma11}%
  \BibitemOpen
  \bibfield  {author} {\bibinfo {author} {\bibfnamefont {S.}~\bibnamefont
  {Das~Sarma}}, \bibinfo {author} {\bibfnamefont {S.}~\bibnamefont {Adam}},
  \bibinfo {author} {\bibfnamefont {E.~H.}\ \bibnamefont {Hwang}}, \ and\
  \bibinfo {author} {\bibfnamefont {E.}~\bibnamefont {Rossi}},\ }\href
  {http://link.aps.org/doi/10.1103/RevModPhys.83.407} {\bibfield  {journal}
  {\bibinfo  {journal} {Rev. Mod. Phys.}\ }\textbf {\bibinfo {volume} {83}},\
  \bibinfo {pages} {407} (\bibinfo {year} {2011})}\BibitemShut {NoStop}%
\bibitem [{\citenamefont {de~Juan}, \citenamefont {Cortijo},\ and\
  \citenamefont {Vozmediano}(2007)}]{Juan07}%
  \BibitemOpen
  \bibfield  {author} {\bibinfo {author} {\bibfnamefont {F.}~\bibnamefont
  {de~Juan}}, \bibinfo {author} {\bibfnamefont {A.}~\bibnamefont {Cortijo}}, \
  and\ \bibinfo {author} {\bibfnamefont {M.~A.~H.}\ \bibnamefont
  {Vozmediano}},\ }\href {http://link.aps.org/doi/10.1103/PhysRevB.76.165409}
  {\bibfield  {journal} {\bibinfo  {journal} {Phys. Rev. B}\ }\textbf {\bibinfo
  {volume} {76}},\ \bibinfo {pages} {165409} (\bibinfo {year}
  {2007})}\BibitemShut {NoStop}%
\bibitem [{\citenamefont {Wu}\ \emph {et~al.}(2007)\citenamefont {Wu},
  \citenamefont {Li}, \citenamefont {Song}, \citenamefont {Berger},\ and\
  \citenamefont {de~Heer}}]{Wu07}%
  \BibitemOpen
  \bibfield  {author} {\bibinfo {author} {\bibfnamefont {X.}~\bibnamefont
  {Wu}}, \bibinfo {author} {\bibfnamefont {X.}~\bibnamefont {Li}}, \bibinfo
  {author} {\bibfnamefont {Z.}~\bibnamefont {Song}}, \bibinfo {author}
  {\bibfnamefont {C.}~\bibnamefont {Berger}}, \ and\ \bibinfo {author}
  {\bibfnamefont {W.~A.}\ \bibnamefont {de~Heer}},\ }\href
  {http://link.aps.org/doi/10.1103/PhysRevLett.98.136801} {\bibfield  {journal}
  {\bibinfo  {journal} {Phys. Rev. Lett.}\ }\textbf {\bibinfo {volume} {98}},\
  \bibinfo {pages} {136801} (\bibinfo {year} {2007})}\BibitemShut {NoStop}%
\bibitem [{\citenamefont {Tikhonenko}\ \emph {et~al.}(2008)\citenamefont
  {Tikhonenko}, \citenamefont {Horsell}, \citenamefont {Gorbachev},\ and\
  \citenamefont {Savchenko}}]{Tikhonenko08}%
  \BibitemOpen
  \bibfield  {author} {\bibinfo {author} {\bibfnamefont {F.~V.}\ \bibnamefont
  {Tikhonenko}}, \bibinfo {author} {\bibfnamefont {D.~W.}\ \bibnamefont
  {Horsell}}, \bibinfo {author} {\bibfnamefont {R.~V.}\ \bibnamefont
  {Gorbachev}}, \ and\ \bibinfo {author} {\bibfnamefont {A.~K.}\ \bibnamefont
  {Savchenko}},\ }\href
  {http://link.aps.org/doi/10.1103/PhysRevLett.100.056802} {\bibfield
  {journal} {\bibinfo  {journal} {Phys. Rev. Lett.}\ }\textbf {\bibinfo
  {volume} {100}},\ \bibinfo {pages} {056802} (\bibinfo {year}
  {2008})}\BibitemShut {NoStop}%
\bibitem [{\citenamefont {Ki}\ \emph {et~al.}(2008)\citenamefont {Ki},
  \citenamefont {Jeong}, \citenamefont {Choi}, \citenamefont {Lee},\ and\
  \citenamefont {Park}}]{Ki08}%
  \BibitemOpen
  \bibfield  {author} {\bibinfo {author} {\bibfnamefont {D.-K.}\ \bibnamefont
  {Ki}}, \bibinfo {author} {\bibfnamefont {D.}~\bibnamefont {Jeong}}, \bibinfo
  {author} {\bibfnamefont {J.-H.}\ \bibnamefont {Choi}}, \bibinfo {author}
  {\bibfnamefont {H.-J.}\ \bibnamefont {Lee}}, \ and\ \bibinfo {author}
  {\bibfnamefont {K.-S.}\ \bibnamefont {Park}},\ }\href
  {http://link.aps.org/doi/10.1103/PhysRevB.78.125409} {\bibfield  {journal}
  {\bibinfo  {journal} {Phys. Rev. B}\ }\textbf {\bibinfo {volume} {78}},\
  \bibinfo {pages} {125409} (\bibinfo {year} {2008})}\BibitemShut {NoStop}%
\bibitem [{\citenamefont {Lara-Avila}\ \emph
  {et~al.}(2011{\natexlab{b}})\citenamefont {Lara-Avila}, \citenamefont
  {Tzalenchuk}, \citenamefont {Kubatkin}, \citenamefont {Yakimova},
  \citenamefont {Janssen}, \citenamefont {Cedergren}, \citenamefont
  {Bergsten},\ and\ \citenamefont {Fal'ko}}]{LaraAvila11}%
  \BibitemOpen
  \bibfield  {author} {\bibinfo {author} {\bibfnamefont {S.}~\bibnamefont
  {Lara-Avila}}, \bibinfo {author} {\bibfnamefont {A.}~\bibnamefont
  {Tzalenchuk}}, \bibinfo {author} {\bibfnamefont {S.}~\bibnamefont
  {Kubatkin}}, \bibinfo {author} {\bibfnamefont {R.}~\bibnamefont {Yakimova}},
  \bibinfo {author} {\bibfnamefont {T.~J. B.~M.}\ \bibnamefont {Janssen}},
  \bibinfo {author} {\bibfnamefont {K.}~\bibnamefont {Cedergren}}, \bibinfo
  {author} {\bibfnamefont {T.}~\bibnamefont {Bergsten}}, \ and\ \bibinfo
  {author} {\bibfnamefont {V.}~\bibnamefont {Fal'ko}},\ }\href
  {http://link.aps.org/doi/10.1103/PhysRevLett.107.166602} {\bibfield
  {journal} {\bibinfo  {journal} {Phys. Rev. Lett.}\ }\textbf {\bibinfo
  {volume} {107}},\ \bibinfo {pages} {166602} (\bibinfo {year}
  {2011}{\natexlab{b}})}\BibitemShut {NoStop}%
\bibitem [{\citenamefont {Huang}\ \emph {et~al.}(2015)\citenamefont {Huang},
  \citenamefont {Alexander-Webber}, \citenamefont {Baker}, \citenamefont
  {Janssen}, \citenamefont {Tzalenchuk}, \citenamefont {Antonov}, \citenamefont
  {Yager}, \citenamefont {Lara-Avila}, \citenamefont {Kubatkin}, \citenamefont
  {Yakimova},\ and\ \citenamefont {Nicholas}}]{Huang15}%
  \BibitemOpen
  \bibfield  {author} {\bibinfo {author} {\bibfnamefont {J.}~\bibnamefont
  {Huang}}, \bibinfo {author} {\bibfnamefont {J.~A.}\ \bibnamefont
  {Alexander-Webber}}, \bibinfo {author} {\bibfnamefont {A.~M.~R.}\
  \bibnamefont {Baker}}, \bibinfo {author} {\bibfnamefont {T.~J. B.~M.}\
  \bibnamefont {Janssen}}, \bibinfo {author} {\bibfnamefont {A.}~\bibnamefont
  {Tzalenchuk}}, \bibinfo {author} {\bibfnamefont {V.}~\bibnamefont {Antonov}},
  \bibinfo {author} {\bibfnamefont {T.}~\bibnamefont {Yager}}, \bibinfo
  {author} {\bibfnamefont {S.}~\bibnamefont {Lara-Avila}}, \bibinfo {author}
  {\bibfnamefont {S.}~\bibnamefont {Kubatkin}}, \bibinfo {author}
  {\bibfnamefont {R.}~\bibnamefont {Yakimova}}, \ and\ \bibinfo {author}
  {\bibfnamefont {R.~J.}\ \bibnamefont {Nicholas}},\ }\href
  {http://link.aps.org/doi/10.1103/PhysRevB.92.075407} {\bibfield  {journal}
  {\bibinfo  {journal} {Phys. Rev. B}\ }\textbf {\bibinfo {volume} {92}},\
  \bibinfo {pages} {075407} (\bibinfo {year} {2015})}\BibitemShut {NoStop}%
\bibitem [{\citenamefont {Tanabe}\ \emph {et~al.}(2011)\citenamefont {Tanabe},
  \citenamefont {Sekine}, \citenamefont {Kageshima}, \citenamefont {Nagase},\
  and\ \citenamefont {Hibino}}]{Tanabe11}%
  \BibitemOpen
  \bibfield  {author} {\bibinfo {author} {\bibfnamefont {S.}~\bibnamefont
  {Tanabe}}, \bibinfo {author} {\bibfnamefont {Y.}~\bibnamefont {Sekine}},
  \bibinfo {author} {\bibfnamefont {H.}~\bibnamefont {Kageshima}}, \bibinfo
  {author} {\bibfnamefont {M.}~\bibnamefont {Nagase}}, \ and\ \bibinfo {author}
  {\bibfnamefont {H.}~\bibnamefont {Hibino}},\ }\href
  {http://link.aps.org/doi/10.1103/PhysRevB.84.115458} {\bibfield  {journal}
  {\bibinfo  {journal} {Phys. Rev. B}\ }\textbf {\bibinfo {volume} {84}},\
  \bibinfo {pages} {115458} (\bibinfo {year} {2011})}\BibitemShut {NoStop}%
\bibitem [{\citenamefont {Giesbers}, \citenamefont {Proch\'azka},\ and\
  \citenamefont {Flipse}(2013)}]{Giesbers13}%
  \BibitemOpen
  \bibfield  {author} {\bibinfo {author} {\bibfnamefont {A.~J.~M.}\
  \bibnamefont {Giesbers}}, \bibinfo {author} {\bibfnamefont {P.}~\bibnamefont
  {Proch\'azka}}, \ and\ \bibinfo {author} {\bibfnamefont {C.~F.~J.}\
  \bibnamefont {Flipse}},\ }\href
  {http://link.aps.org/doi/10.1103/PhysRevB.87.195405} {\bibfield  {journal}
  {\bibinfo  {journal} {Phys. Rev. B}\ }\textbf {\bibinfo {volume} {87}},\
  \bibinfo {pages} {195405} (\bibinfo {year} {2013})}\BibitemShut {NoStop}%
\bibitem [{\citenamefont {Yu}\ \emph {et~al.}(2013)\citenamefont {Yu},
  \citenamefont {Li}, \citenamefont {Liu}, \citenamefont {Dun}, \citenamefont
  {He}, \citenamefont {Zhang}, \citenamefont {Cai},\ and\ \citenamefont
  {Feng}}]{Yu13}%
  \BibitemOpen
  \bibfield  {author} {\bibinfo {author} {\bibfnamefont {C.}~\bibnamefont
  {Yu}}, \bibinfo {author} {\bibfnamefont {J.}~\bibnamefont {Li}}, \bibinfo
  {author} {\bibfnamefont {Q.~B.}\ \bibnamefont {Liu}}, \bibinfo {author}
  {\bibfnamefont {S.~B.}\ \bibnamefont {Dun}}, \bibinfo {author} {\bibfnamefont
  {Z.~Z.}\ \bibnamefont {He}}, \bibinfo {author} {\bibfnamefont {X.~W.}\
  \bibnamefont {Zhang}}, \bibinfo {author} {\bibfnamefont {S.~J.}\ \bibnamefont
  {Cai}}, \ and\ \bibinfo {author} {\bibfnamefont {Z.~H.}\ \bibnamefont
  {Feng}},\ }\href
  {http://scitation.aip.org/content/aip/journal/apl/102/1/10.1063/1.4773568}
  {\bibfield  {journal} {\bibinfo  {journal} {Applied Physics Letters}\
  }\textbf {\bibinfo {volume} {102}},\ \bibinfo {eid} {013107} (\bibinfo {year}
  {2013})}\BibitemShut {NoStop}%
\bibitem [{\citenamefont {Kalugin}\ \emph {et~al.}(2011)\citenamefont
  {Kalugin}, \citenamefont {Jing}, \citenamefont {Bao}, \citenamefont {Wickey},
  \citenamefont {Del~Barga}, \citenamefont {Ovezmyradov}, \citenamefont
  {Shaner},\ and\ \citenamefont {Lau}}]{Kalugin11}%
  \BibitemOpen
  \bibfield  {author} {\bibinfo {author} {\bibfnamefont {N.~G.}\ \bibnamefont
  {Kalugin}}, \bibinfo {author} {\bibfnamefont {L.}~\bibnamefont {Jing}},
  \bibinfo {author} {\bibfnamefont {W.}~\bibnamefont {Bao}}, \bibinfo {author}
  {\bibfnamefont {L.}~\bibnamefont {Wickey}}, \bibinfo {author} {\bibfnamefont
  {C.}~\bibnamefont {Del~Barga}}, \bibinfo {author} {\bibfnamefont
  {M.}~\bibnamefont {Ovezmyradov}}, \bibinfo {author} {\bibfnamefont {E.~A.}\
  \bibnamefont {Shaner}}, \ and\ \bibinfo {author} {\bibfnamefont {C.~N.}\
  \bibnamefont {Lau}},\ }\href
  {http://scitation.aip.org/content/aip/journal/apl/99/1/10.1063/1.3609320}
  {\bibfield  {journal} {\bibinfo  {journal} {Applied Physics Letters}\
  }\textbf {\bibinfo {volume} {99}},\ \bibinfo {eid} {013504} (\bibinfo {year}
  {2011})}\BibitemShut {NoStop}%
\bibitem [{\citenamefont {Kalugin}\ \emph {et~al.}(2002)\citenamefont
  {Kalugin}, \citenamefont {Vasilyev}, \citenamefont {Suchalkin}, \citenamefont
  {Nachtwei}, \citenamefont {Sagol},\ and\ \citenamefont {Eberl}}]{Vas2002}%
  \BibitemOpen
  \bibfield  {author} {\bibinfo {author} {\bibfnamefont {N.~G.}\ \bibnamefont
  {Kalugin}}, \bibinfo {author} {\bibfnamefont {Y.~B.}\ \bibnamefont
  {Vasilyev}}, \bibinfo {author} {\bibfnamefont {S.~D.}\ \bibnamefont
  {Suchalkin}}, \bibinfo {author} {\bibfnamefont {G.}~\bibnamefont {Nachtwei}},
  \bibinfo {author} {\bibfnamefont {B.~E.}\ \bibnamefont {Sagol}}, \ and\
  \bibinfo {author} {\bibfnamefont {K.}~\bibnamefont {Eberl}},\ }\href
  {http://link.aps.org/doi/10.1103/PhysRevB.66.085308} {\bibfield  {journal}
  {\bibinfo  {journal} {Phys. Rev. B}\ }\textbf {\bibinfo {volume} {66}},\
  \bibinfo {pages} {085308} (\bibinfo {year} {2002})}\BibitemShut {NoStop}%
\end{thebibliography}%


\end{document}